\documentclass[conference]{IEEEtran}

\usepackage{graphicx}
\usepackage{tabularx}

\usepackage[T1]{fontenc}
\usepackage[utf8]{inputenc}

\usepackage{csquotes}
\usepackage{nicefrac}
\usepackage{afterpage}
\usepackage{needspace}
\usepackage[textsize=scriptsize]{todonotes}
\usepackage{booktabs}
\usepackage{tikz}
\usepackage{quantikz}
\usetikzlibrary{arrows,positioning} 
\usetikzlibrary{shapes.geometric, positioning}
\usepackage{flushend}
\usepackage{soul}
\usepackage{amssymb}
\usepackage{csvsimple}
\usepackage{amsthm}
\usepackage{lipsum}
\usepackage{varwidth}
\usepackage{utfsym}
\usepackage{fontawesome5}
\usepackage{multicol}
\usepackage{listings}
\usepackage{adjustbox}
\usepackage{float}
\usepackage{placeins}

\definecolor{darkgreen}{RGB}{0, 150, 0}

\newfloat{lstfloat}{htbp}{lop}
\floatname{lstfloat}{\footnotesize Listing}

\robustify\bfseries

\makeatletter
\let\MYcaption\@makecaption
\makeatother
\usepackage[font=footnotesize]{subcaption}
\makeatletter
\let\@makecaption\MYcaption
\makeatother

\usepackage{yquant}
\usepackage{pgfplots}

\usepackage[style=ieee, maxnames=6, minnames=1, date=year, doi=false,isbn=false,backend=biber]{biblatex}
\usepackage[detect-all=true]{siunitx}
\usepackage{etoolbox}
\robustify\bfseries

\usetikzlibrary{fit, quotes}
\usetikzlibrary{shapes.geometric, positioning}
\usepackage{bbm}

\newtheorem{example}{Example}

\usetikzlibrary{automata,arrows}

\AtEveryBibitem{
  \clearname{editor}%
  \clearfield{series}%
  \clearfield{isbn}%
  \clearfield{issn}%
  \clearfield{volume}%
  \clearfield{number}%
  \clearfield{pages}%
  \clearfield{doi}%
}

\renewcommand{\baselinestretch}{1.0}
\lstdefinelanguage{mqt-qasm}{
  keywords={include, qreg, creg, gate, measure, reset, barrier, if, opaque},
  keywordstyle=\color{blue}\bfseries,
  ndkeywords={x, h, cccx, ccx, cx, z, oracle, diffusion},
  ndkeywordstyle=\color{darkgray}\bfseries,
  identifierstyle=\color{black},
  sensitive=false,
  comment=[l]{//},
  morecomment=[s]{/*}{*/},
  commentstyle=\color{darkgreen}\ttfamily,
  stringstyle=\color{red}\ttfamily,
  morestring=[b]",
  morestring=[b]',
  basicstyle=\ttfamily\footnotesize,
  numbers=left,
  numberstyle=\tiny\color{gray},
  stepnumber=1,
  numbersep=5pt,
  showspaces=false,
  showstringspaces=false,
  showtabs=false,
  tabsize=2,
  breaklines=true,
  breakatwhitespace=false,
  captionpos=b,
  frame=single,
  alsoletter=-,
  lineskip=0pt,
}

\lstdefinestyle{qasm}{
    commentstyle=\color{darkgreen},
    keywordstyle=\color{blue},
    numberstyle=\tiny\color{gray},
    stringstyle=\color{red},
    basicstyle=\ttfamily\footnotesize,
    breakatwhitespace=false,         %
    breaklines=true,                 %
    captionpos=t,                    %
    keepspaces=true,                 %
    numbers=left,                    %
    numbersep=5pt,                   %
    showspaces=false,                %
    showstringspaces=false,          %
    showtabs=false,                  %
    tabsize=2,                       %
    frame=single,                    %
}
\newcommand{\revisedOld}[1]{#1}

\newif\ifdoubleblind
\doubleblindfalse

\newcommand{\redacted}[1]{\ifdoubleblind[redacted for double-blind review]\else#1\fi}
\newcommand{\redactedInstead}[2]{\ifdoubleblind#2\else#1\fi}

\redactedInstead{\renewcommand{\baselinestretch}{0.95}}{\renewcommand{\baselinestretch}{0.975}}

\usepackage{hyperref}

\hypersetup{
	pdftitle={Automatically Refining Assertions for Efficient Debugging of Quantum Programs},   %
	pdfsubject={Design Automation Conference 2025}, %
	pdfauthor={\redactedInstead{Damian Rovara, Lukas Burgholzer, Robert Wille}{}}
}

\begin{document}

\title{Automatically Refining Assertions for \\ Efficient Debugging of Quantum Programs}

\author{
  \redactedInstead{
	\IEEEauthorblockN{Damian Rovara\IEEEauthorrefmark{1}\hspace*{1.5cm}Lukas Burgholzer\IEEEauthorrefmark{1}\hspace*{1.5cm}Robert Wille\IEEEauthorrefmark{1}\IEEEauthorrefmark{2}}
	\IEEEauthorblockA{\IEEEauthorrefmark{1}Chair for Design Automation, Technical University of Munich, Germany}
	\IEEEauthorblockA{\IEEEauthorrefmark{2}Software Competence Center Hagenberg GmbH (SCCH), Austria}
	\IEEEauthorblockA{\href{mailto:damian.rovara@tum.de}{damian.rovara@tum.de}\hspace{1.5cm}\href{mailto:lukas.burgholzer@tum.de}{lukas.burgholzer@tum.de}\hspace{1.5cm} \href{mailto:robert.wille@tum.de}{robert.wille@tum.de}\\
	\url{https://www.cda.cit.tum.de/research/quantum}
	}
  }{\vspace{-0.0cm}}
}

\maketitle

\begin{abstract}
  As new advancements in the field of quantum computing lead to the development of increasingly complex programs,
  approaches to validate and debug these programs are becoming more important. To this end, methods employed in classical
  debugging, such as assertions for testing specific properties of a program's state, have been adapted for quantum programs.
  However, to efficiently debug quantum programs, it is key to properly place these assertions. This usually requires a deep understanding
  of the program's underlying mathematical properties, constituting a time-consuming manual task for developers.
  To address this problem, this work proposes methods for \emph{automatically} refining assertions in quantum programs by moving them to more favorable positions
  in the program or by placing new assertions that help to further narrow down potential error locations.
  This allows developers to take advantage of rich and expressive assertions that greatly improve the debugging experience without requiring them to 
  place these assertions manually in an otherwise tedious manner. 
  An open-source implementation of the proposed methods is available at \redacted{\href{https://github.com/cda-tum/mqt-debugger}{https://github.com/cda-tum/mqt-debugger}}.
\end{abstract}

\vspace{-0.1cm}
\section{Introduction}\label{sec:introduction}

Assertions are a powerful tool commonly employed to test and verify (classical) programs~\cite{rosenblum1995, clarke2006, taromirad2024}.
They allow developers to specify properties that should hold at a specific time during the execution of a program,
which are then checked. If an assertion fails, this typically indicates an error in the program, as
the program's state differs from what was expected.

With the recent growth of quantum computing, it is becoming more important to develop efficient debugging strategies
for quantum programs~\cite{dimatteo2024, huang2019a, li2014, miranskyy2020, miranskyy2021, metwalli2022, chen2023, garciadelabarrera2023, zhao2023, metwalli2024}. Therefore, assertions have similarly started to be employed in quantum \mbox{programs \cite{liu2019, haner2020, yu2021, ying2022, bauer-marquart2023, huang2019, li2020, liu2020, liu2021, li2022, witharana2023}}.
In many cases, they can drastically reduce the workload of debugging quantum programs, an otherwise tedious and \mbox{time-consuming}
task due to the exponential size of the state space of quantum programs, which often requires developers to compute and track a large number of
amplitudes. By placing assertions within a program, developers can narrow down the search space for errors: An assertion that fails
indicates that the error must be located in some part of the program that has already been executed.

However, the amount of information that can be inferred from an assertion failure strongly depends on the assertion's quality. While very
general assertions that cover a large part of the program's state space have a high likelihood of detecting the presence of an error, it
is often difficult to infer what exactly caused the error. On the other hand, assertions that are too specific may not detect errors at all.
Furthermore, the placement of assertions within a program is crucial: If assertions are placed too late in the program, the information that can be
inferred from them can be limited. Assertions that fail at the end of a program, for example, can make it difficult to
determine whether the error occurred at the earlier or later parts of the program.

To efficiently debug a quantum program, it is therefore crucial to select assertions that maximize the information that can be inferred from their failures.
However, this once again suffers from similar challenges as discussed above: The complex nature of quantum programs makes it difficult
to determine the exact state of the program at any given time, often requiring a deep understanding of its mathematical properties. 
Furthermore, the complex interactions between qubits in quantum programs further obfuscate this state even to experienced developers. 
This, in turn, makes it difficult to determine what assertions
should be placed where in a quantum program.

To address this issue, automated methods to assist developers in the placement of assertions in quantum programs are required.
While such methods already exist in classical debugging, such as \emph{GoldMine}~\cite{vasudevan2010}
and Daikon~\cite{ernst2007}, two tools to inspect classical programs and automatically generate assertions from the corresponding
findings, \revisedOld{applying similar methods to quantum programs is often not sufficient to generate high-quality assertions, as discussed later in~\autoref{sec:background}.}

To resolve this limitation, this work proposes methods to automatically refine assertions in quantum programs. These methods allow
developers to define the intent of a quantum program through a set of initial assertions, which are then refined to reduce the
required debugging workload. By moving assertions to earlier locations in the program, these methods can bring assertions closer to 
the detected error, making it easier to find the exact problem cause.
Furthermore, by adding new assertions that inspect the program's state in more detail, the search space for errors can be reduced even further.
All proposed methods are available as an open-source implementation at \redacted{\href{https://github.com/cda-tum/mqt-debugger}{https://github.com/cda-tum/mqt-debugger}}.

Evaluations confirm that the proposed methods can significantly improve the quality of assertions in many quantum programs,
reducing the number of instructions to be inspected between failing assertions and the error they detect by up to $61\%$. Additionally, these evaluation
results show that the proposed methods can be applied to a wide range of quantum programs, including several well-known quantum algorithms.

The remainder of this work is structured as follows: 
\autoref{sec:background} provides an overview of the usage of assertions in quantum programs. \autoref{sec:motivation} then illustrates the
challenges of manually placing assertions in quantum programs, motivating the need for automated methods and suggesting a general idea
to circumvent these challenges. \autoref{sec:methods} then introduces the proposed methods for assertion refinement in quantum programs,
focusing on the \emph{movement} and \emph{addition} of assertions. \autoref{sec:evaluation} evaluates the proposed methods on a set of quantum programs,
showcasing that the proposed methods can significantly improve the quality of assertions in these programs. Finally, \autoref{sec:conclusion} concludes
the work.

\section{Background}\label{sec:background}

While assertions have been widely used in classical programming, recent work has
also adapted them for quantum programs~\cite{liu2019, haner2020, yu2021, ying2022, bauer-marquart2023, huang2019, li2020, liu2020, liu2021, li2022, witharana2023}.
However, limitations of quantum mechanics make the implementation of assertions on real quantum devices more challenging. As measurements of the program's state during
execution may cause side effects, assertions typically cannot test for arbitrary states in a straightforward manner.

Because of this, several approaches have been proposed to circumvent these limitations or to reduce their impact. 
Projection-based assertions, as proposed by Li et al.~\cite{li2020}, are one such approach, allowing specific states to be evaluated 
through projective measurements. Furthermore, Liu et \mbox{al. \cite{liu2020, liu2021}} have proposed the use of
ancillary qubits to test certain properties of the program. Alternatively, statistical assertions,
as proposed by Huang et al.~\cite{huang2019}, can test the state of programs by sampling it over repeated executions
and comparing the results with expected values.

While, typically, assertions have to be placed manually in quantum programs, different strategies
that aim to aid in this process have already been proposed. Ying et al.~\cite{ying2017} have proposed
a method to automatically generate invariants for programs defined in a quantum-\textbf{while}-language, 
which play an important role in program verification and can be used as assertions. Furthermore,
Witharana et al.~\cite{witharana2023} introduced \emph{quAssert}, a tool for the automatic generation of
quantum assertions based on the static analysis and random sampling of quantum programs. However,
both of these approaches focus on the generation of assertions based on the program's current structure.
While this approach allows for efficient regression testing after changes are applied to the program,
it cannot help in the detection of errors that are already present. This, once again, leads to
a large manual effort, requiring developers to investigate the generated assertions and the program's structure
which is often more complex for quantum programs compared to their classical counterparts.

This work focuses on three representative types of assertions, frequently used in debugging of quantum 
programs~\cite{huang2019, liu2020, witharana2023}, namely
\begin{itemize}
    \item \emph{superposition assertions} checking, whether the given qubits are in a superposition state,
    \item \emph{entanglement assertions} checking, whether the given qubits are in a fully entangled state, and
    \item \emph{equality assertions} checking, whether the given qubits are in a specified state.
\end{itemize}

Focusing on these three types of assertions is sufficient to motivate the problem considered in this work
as well as describing the proposed solutions. The following discussions and proposals
can easily be extended with further types of assertions. These assertion types can be integrated in quantum programs, such as \emph{OpenQASM}~\cite{cross2022openqasm} code,
to test the correctness of the program at any step during execution. An example usage of each of
these assertion types in OpenQASM code is illustrated at the end of \autoref{lst:cccx}.
Their specific implementation is left open, they may be implemented as statistical
assertions on real quantum devices or tested on classically-simulated quantum computers.

\section{Motivation and General Idea}\label{sec:motivation}
While the use of assertions has proven to be an effective method to
find errors in quantum programs, placing them manually can be a time-consuming
task for developers. This section illustrates
the challenges and limitations of manually placing assertions. It then devises a general idea to \emph{automatically} refine
assertions in quantum programs to assist developers in this task.

\subsection{Considered Problem}

The quality of assertions employed when debugging quantum programs greatly affects the efficiency
at which underlying errors can be found. Well-placed assertions can
significantly narrow the potential error location down to a small region in the
program being debugged. However, as quantum programs grow in size and complexity, 
determining how to place assertions well becomes increasingly difficult.

\begin{example}
\label{ex:motivation}
Consider \autoref{lst:cccx} that shows a quantum program applying a \mbox{three-controlled \texttt{X} gate
~(\texttt{CCCX})} to the state~$|\mathord{+}\mathord{+}\mathord{+}\rangle$
using two ancillary qubits followed by a single-qubit \texttt{Z} gate.
To test for its correctness, several
assertions are employed at the end of the program. The assertion on Line~21 checks whether the ancilla qubits have
been uncomputed correctly by requiring their state to be equal to $|00\rangle$. The assertion on Line~22 checks whether
\texttt{target[0]} is in a superposition after applying all operations. Finally, the assertion on Line~23 checks whether
the qubits~\texttt{q[0]} and~\texttt{target[0]} are in an entangled state.

However, this program has an error: On Line~10, the order of \texttt{anc[0]} and \texttt{anc[1]} is 
swapped---causing the circuit to apply incorrect operations. The assertions at the end of the program are able to
correctly determine that an error has occurred: As the ancilla qubits are not prepared correctly, the uncomputation
will not leave them in the \mbox{$|00\rangle$-state}, causing the assertion on Line~21 to fail. Furthermore, as \texttt{anc[1]}
is in a purely \mbox{$|0\rangle$-state} when executing Line~10, \texttt{target[0]} will remain unchanged and will not be in a superposition
or entangled with \texttt{q[0]}, causing the assertions on Line~22 and 23 to fail. Therefore, running this
program and checking the assertions will result in all three of the assertions failing.
\end{example}

While assertions such as those in \autoref{ex:motivation} indeed allow
developers to identify that there is \emph{something} wrong 
with the program, it is far from obvious \emph{what exactly} caused the issue.
This is because by the time the assertions are reached at the end of the execution, the entire program has
already been executed. Therefore, the only possible inference that can be made is that the error is located somewhere
in the program. It is still left to the developer to find
the exact location of those errors, requiring the entire program
to be investigated manually in a time-consuming process.

\revisedOld{Instead, placing assertions at earlier locations in the program
may greatly improve the information that can be inferred from
their failures. However, this comes with several challenges. Due to
the complex interactions between individual qubits in quantum programs,
determining the earliest location at which a given assertion can be placed
requires deep understanding
of the program's underlying mathematical properties.
Furthermore, in many cases, complex assertions can be broken down into
simpler parts that provide more information for debugging. Therefore, the
manual refinement of assertions is a time-consuming task, raising the need
for automated methods to assist developers in this process.}

\begin{lstfloat}[t]
\begin{lstlisting}[language=mqt-qasm]
qreg q[3];
qreg anc[2];
qreg target[1];

// prepare |+++>
h q;

// Apply CCCX gate
ccx q[0], q[1], anc[0];
ccx q[2], anc[1], anc[0];
cx anc[1], target[0];

// Uncompute
ccx q[2], anc[0], anc[1];
ccx q[0], q[1], anc[0];

// Apply Pauli-Z gate
z target[0]

// Check
assert-eq anc[0], anc[1] { 1, 0, 0, 0 }
assert-sup target;
assert-ent q[0], target[0];
\end{lstlisting}
\vspace{-0.45cm}
\caption{\footnotesize An incorrect implementation of a quantum circuit that applies a \texttt{CCCX} gate and a \texttt{Z} gate to the qubit \texttt{target[0]}. 
It uses assertions to check for correctness.\vspace{-0.6cm}}
\label{lst:cccx}
\end{lstfloat}
\refstepcounter{lstlisting}

\subsection{General Idea}
\revisedOld{To assist developers in using assertions in a way that reduces the required manual debugging effort, 
we propose multiple methods to \emph{automatically} refine existing assertions. First,}
we attempt to move existing assertions up the program as far as possible. 
Typically, placing assertions earlier in the code helps reduce the number of error candidates, as
the number of possibly incorrect operations also decreases. This can be done by defining \emph{commutation rules}
for the different types of assertions. This way, we can iteratively
try to exchange each individual assertion with its predecessor instruction, as long as they commute with each other.
Depending on the structure of the program, this approach can move a large number of assertions efficiently,
especially assertions that are placed at the very end of the program, as is often the case in classical testing and debugging.

\begin{example}
    \label{ex:movement}
    Returning to \autoref{ex:motivation} and \autoref{lst:cccx}, commutation rules can be employed to move all
    assertions at the end of the program to earlier positions. First, the \texttt{Z} gate applied to \texttt{target[0]} on 
    Line~18 clearly cannot influence the correctness of the equality
    assertion on Line~21, as it only considers the qubits~\texttt{anc[0]} and \texttt{anc[1]}. Therefore, the assertion can
    be moved above this gate without further concerns. The superposition assertion on Line~22 can be moved in a similar
    manner. While it considers the same qubit that is acted on by Line~18, a single \texttt{Z} gate cannot influence the
    correctness of this assertion, as it does not change whether a qubit is in a superposition state. Furthermore, this
    assertion can be moved further up to Line~12, as all instructions below that do not directly involve \texttt{target[0]}.
    Finally, similar logic can be applied to the entanglement assertion on Line~23: A single \texttt{Z} gate cannot
    influence entanglement between two qubits. Therefore, this assertion can also be moved
    to Line~16.

    As all moved assertions still fail, since commutativity assures that the assertion results remain the same even after being moved,
    this allows developers to reduce the number of lines that have to be investigated for errors from seven in the original case to
    four through the superposition assertion moved to Line~12.
\end{example}

These commutation rules allow moving assertions to
a more favorable position. This is shown particularly for the
superposition assertion from the previous example, where the
assertion was moved halfway through the whole program. 
Moving assertions up using these commutation rules is already very effective.
In fact, as discussed later in \autoref{sec:evaluation}, moving assertions can often drastically
reduce the number of instructions that need to be considered during debugging by bringing them closer to the original
error position. However, more
potential remains.

In fact, the debugging experience can be optimized even further by
not only \emph{moving} assertions but also by automatically \emph{adding} new assertions based on the existing ones.
Through static program analysis, certain preconditions of existing assertions
can be derived and turned into additional assertions.

\begin{example}
    \autoref{ex:movement} has shown how the assertion on Line~23 in \autoref{lst:cccx} can be moved up to Line~16. However,
    analyzing the interactions between the individual qubits in this program and collecting them in an \emph{interaction graph}
    shows that the qubits \texttt{q[0]} and \texttt{target[0]} are indirectly connected through the qubit \texttt{anc[1]}.
    Therefore, we can infer that \texttt{target[0]} and \texttt{anc[1]} must have been entangled after their last interaction
    as well. This allows us to place an additional assertion, \texttt{assert-ent anc[1], target} on Line~12.
    
    After executing the program and checking the assertions again, the new assertion also fails, as \texttt{anc[1]} and \texttt{target[0]}
    are not entangled at this point. Therefore, we can conclude that the error must have occurred before Line~12 already, reducing
    the number of lines that have to be investigated for errors even further from seven to four, \revisedOld{similar to the results
    obtained by moving the superposition assertion in \autoref{ex:movement}}.
\end{example}

Naturally, the commutation rules
discussed above may once again be employed to these newly added assertions to move them
to better positions.

For both of these approaches, a set of initial assertions is substantially refined---allowing for a much better debugging experience.
This results in a scenario in which the initial assertions still can be provided at a very general level and represent the ground truth for the program's expected behavior as intended by the developer. 
Then, the above methods automatically refine them by determining their optimal
locations and defining additional, more detailed assertions when possible. \revisedOld{In some cases, larger assertions
may even be replaced completely by new, more detailed assertions.}

\section{Proposed Methods}\label{sec:methods}

Based on the general ideas described above, this section describes the respective techniques in more detail.
To this end, we first discuss how assertions can be moved to earlier positions and, then, 
define strategies to add further assertions. \revisedOld{All analyses and operations performed by the proposed methods can be employed statically, without requiring the
quantum programs to be executed.}

\subsection{Moving Assertions}\label{sec:moving}

To increase the amount of information that can be deduced from failing assertions and, therefore,
minimize the manual workload for debugging quantum programs, we attempt to
move assertions to earlier locations in the program. As an assertion can only be moved above
instructions that cannot influence its outcome, this method defines a set of \emph{commutation rules} for all assertion types.
Each assertion is iteratively compared with its predecessor instruction. If any commutation rule
states that the assertion and its predecessor instruction commute, the assertion is moved above its predecessor.
This process is repeated, until an instruction is reached that does not commute with the assertion.

Due to the different scopes of the assertion types, commutation rules may differ in strictness, depending
on what type of assertion they apply to. In this work, we propose five commutation rules 
to be employed to move assertions. These rules can either apply generally to any type of assertion
or be related to specific assertion types. More precisely:

\begin{enumerate}
    \item \label{rule:non-func} \emph{Non-Functional Instructions}: Instructions representing operations that do not
    affect the current state of the program---gate definitions, register definitions, and barrier operations---always
    commute with \emph{any type of assertion}.
    \item \label{rule:disjunct} \emph{Disjunct Target Instructions}: Instructions that perform non-measurement operations on qubits
    that are not part of an assertion's set of target qubits commute with \emph{any type of assertion}.
    \item \label{rule:diagonal} \emph{(Anti-)Diagonal Instructions}: Instructions with operator matrices that can be represented as diagonal
    or \mbox{anti-diagonal} matrices, such as the Pauli operations and global phase operations commute with \emph{superposition
    assertions}.
    \item \label{rule:single} \emph{Single-Qubit Instructions}: Any non-measurement instruction that is only applied to a single qubit always commutes
    with \emph{entanglement assertions}.
    \item \label{rule:measurement} \emph{Measurement Instructions}: Measurement and reset operations lead to a collapse of the state of entangled qubits.
    As it is not possible to determine what qubits are entangled with each other at static time, especially inside custom gate
    definitions, any instruction that requires a measurement does not commute with \emph{any type of assertion}.
\end{enumerate}

\begin{figure}
    \centering
    \includegraphics[width=\columnwidth]{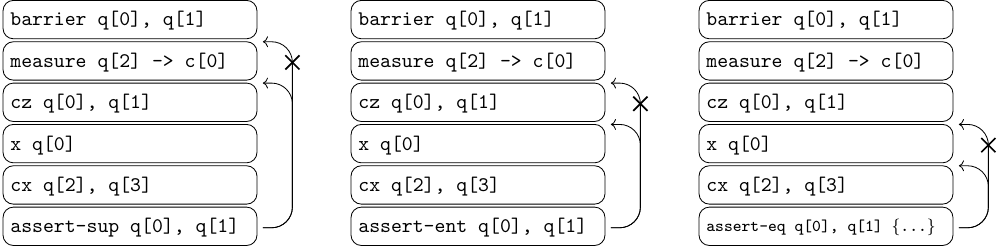}
    \caption{Applications of the commutation rules for the three different types of assertions. The superposition assertion can
    be moved up by three instructions, while the entanglement and equality assertions can be moved up by two and one instructions,
    respectively.}
    \label{fig:commutation}
    \vspace{-0.5cm}
\end{figure}

\begin{example}
\autoref{fig:commutation} shows the application of the commutation rules introduced above for the three types of assertions. 
Using the commutation rules, it can be shown that the \emph{superposition assertion} on the left commutes with the three instructions directly above it: The \texttt{CX}
gate above it does not apply to any of the qubits used in the assertion. Due to Rule~\ref{rule:disjunct},
it therefore commutes with any type of assertion. Furthermore, both the \texttt{X} and the \texttt{CZ} gates can be represented by a diagonal or \mbox{anti-diagonal} operator matrix, thus,
the assertion can also be moved above these instructions, as stated by Rule~\ref{rule:diagonal}.

Similarly, the \emph{entanglement assertion} in the middle of \autoref{fig:commutation} commutes with the \texttt{CX}
gate above due to Rule~\ref{rule:disjunct} once again. While it also commutes with the single-qubit \texttt{X} gate above
due to Rule~\ref{rule:single}, it does not commute with the two-qubit \texttt{CZ} gate. Therefore, the entanglement assertion
can only be moved above its first two proceeding instructions.

Lastly, Rule~\ref{rule:disjunct} once again applies to the \emph{equality assertion} at the right of \autoref{fig:commutation},
but no commutation rules exist that allow it to be moved above the remaining instructions. Because of this, it has to remain
above the \texttt{CX} instruction.

In all three cases, Rule~\ref{rule:measurement} prevents the the assertions from being moved above the measurement instruction,
even if the target qubit \texttt{q[2]} is not considered by the assertion. Because of this, the assertions can also
not be moved up further above the barrier instruction, even though Rule~\ref{rule:non-func} would apply to it, as moving
assertions is stopped, once the first non-commuting instruction is encountered. 
\end{example}

\vspace{-0.3cm}
\subsection{Adding Further Assertions}\label{sec:adding}

The second proposed method to refine assertions is the addition of new assertions. To this end, we propose
two approaches: \emph{Interaction-based} and \emph{State-Separation-based} addition of assertions.
Both approaches aim to add new assertions to the program based on existing assertions initially provided by the developer.
The newly added assertions typically have more limited scopes that
improve the debugging experience by narrowing down potential errors further.

The following provides more details on the proposed methods and discusses in
what situations they can be applied.

\subsubsection*{Interaction-based Addition of Assertions}

\begin{figure}
    \centering
    \includegraphics[width=\columnwidth]{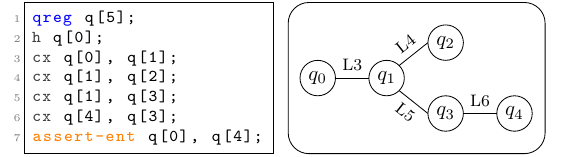}
    \vspace{-0.6cm}
    \caption{A quantum program that generates a 5-qubit GHZ state and the interaction graph constructed by the proposed method. In Line~6,
    control and target have been swapped to induce an error in the program that is detected by the assertion on Line~7. The edges on the
    interaction graph are labled with the line number of the instruction that created them.}
    \label{fig:interaction-generation}
    \vspace{-0.5cm}
\end{figure}

By statically analyzing the interactions of individual qubits in a quantum program, we first generate an
\emph{interaction graph}. \autoref{fig:interaction-generation} illustrates this process for a simple program.
For each gate acting on multiple qubits,
an edge is added to the interaction graph between the qubits used in this instruction. Furthermore, the
interaction graph stores the location of the instruction that created this edge. 

Based on an interaction graph, any entanglement assertion applied to two qubits, \mbox{$q_A$ and $q_B$} can then be refined, by investigating the sub-graph
that connects these two qubits with each other on the interaction graph. If this connection is determined by
a single path \mbox{$\pi = (q_{\pi_1}, q_{\pi_2}, ... , q_{\pi_n})$} such that $q_{\pi_1} = q_A$ and $q_{\pi_n} = q_B$, then new
entanglement assertions can be added, enforcing that $q_{\pi_i}$ and $q_{\pi_{i+1}}$ are entangled with each other for all $i \in [1, n - 1]$. These assertions can
then be inserted after the instructions that added the edge between $q_{\pi_i}$ and $q_{\pi_{i+1}}$ to the interaction graph.

\begin{example}
\autoref{fig:interaction-generation} shows a simple program and the interaction graph generated from it. The assertion on Line~7
tests whether the qubits \texttt{q[0]} and \texttt{q[4]} are in an entangled state. Due to the error introduced in Line~6, where
control and target qubits have been swapped, this assertion will fail. The proposed \emph{Interaction-based} method to add assertions
finds the sub-graph $(q_0, q_1, q_3, q_4)$ that connects both qubits of the assertion with each other.
Based on this sub-graph, it will add the assertions \texttt{assert-ent q[0], q[1]} after Line~3 and \texttt{assert-ent q[0], q[3]} after Line~5.
Running the program again reveals that both of~the new assertions pass, indicating that the error must have occurred between \mbox{Lines~5 and 7},
successfully narrowing down the range of possible errors to the single line that contains the~error.
\end{example}

\subsubsection*{State-Separation-based Addition of Assertions}
\label{sec:state-separation}

The number of target qubits considered by an assertion typically influences, how likely it can be moved to earlier positions using the
methods described above. As an instruction applied to any of these target qubits may possibly prevent the assertion from being moved up further,
it is beneficial to reduce the number of target qubits used by any assertion, as long as this does not influence the general correctness of the assertion.
The \mbox{\emph{State-Separation-based}} method aims to achieve this by splitting existing equality assertions into
smaller assertions, each considering only a subset of the original target qubits.

This is achieved by first investigating the state vector enforced by the assertion and determining all qubits that are separable from it.
For each of these qubits, a new equality assertion is added that only considers this singular qubit. This allows developers to formulate
general equality assertions without needing to consider the aspect of minimizing their scopes, as the proposed method will handle this automatically.

\begin{lstfloat}[t]
\begin{lstlisting}[language=mqt-qasm]
...

cx anc[0], q[0];
(assert-eq anc[0] { 1, 0 })
cx q[1], anc[1];
(assert-eq anc[1] { 1, 0 })
cx q[2], anc[2];
(assert-eq anc[2] { 1, 0 })
assert-eq anc { 1, 0, 0, 0, 0, 0, 0, 0 }
\end{lstlisting}
\vspace{-0.45cm}
\caption{\footnotesize The uncomputation procedure for three qubits in some quantum program. When uncomputing \texttt{anc[0]}, the target and
control qubits are swapped, leading to an error detected by the assertion at the end of the program. The assertions given in
parentheses on \mbox{Lines 4, 6, and 8}
have been added by the \emph{State-Separation-based} method of adding assertions.\\ \vspace{-0.9cm}}
\label{lst:separate}
\begin{tikzpicture}[overlay, remember picture]
    \draw (6.7 + 0.2, 2.05 - 0.02 - 1.7) -- (7.0 + 0.2, 2.05 - 0.02 - 1.7);
    \draw (7.0 + 0.2, 2.05 - 0.02 - 1.7) -- (7.0 + 0.2, 3.6 - 0.02 - 1.7);
    \draw[->] (7.0 + 0.2, 3.6 - 0.02 - 1.7) -- (4.5 + 0.2, 3.6 - 0.02 - 1.7);
    \draw[->] (7.0 + 0.2, 2.98 - 0.02 - 1.7) -- (4.5 + 0.2, 2.98 - 0.02 - 1.7);
    \draw[->] (7.0 + 0.2, 2.35 - 0.02 - 1.7) -- (4.5 + 0.2, 2.35 - 0.02 - 1.7);
\end{tikzpicture}
\end{lstfloat}

\begin{example}
Consider the simple excerpt of a quantum program in \autoref{lst:separate} that performs the uncomputation step for three ancillary qubits
that have been used in previous parts of the program. The assertion at the end of the program checks 
whether the ancillary qubits have been uncomputed correctly and returned to the state $|000\rangle$. However, due
to an error in uncomputing \texttt{anc[0]}, uncomputation is performed incorrectly and the assertion fails. While this
failed assertion indicates an error in the uncomputation procedure, it does not provide any information on which qubit exactly
the error occurred. Due to the \texttt{CX} gate directly above the assertion that involves \texttt{anc[2]}, this assertion cannot
be moved further up.

The \emph{State-Separation-based} method of adding assertions can be employed to assist in this situation. As the state~$|000\rangle$
is separable, three new assertions can be added, each requiring only one of the ancillary qubits to be in the state~$|0\rangle$.
These assertions can then be moved to the positions after the respective \texttt{CX} gates. Running the program again reveals that
the first newly added assertion fails already, indicating that the error occurred during its uncomputation. 
\end{example}

\section{Evaluation}\label{sec:evaluation}

\begin{table*}[t!]
    \caption{The average reduction of lines between the introduced error and the first assertion detecting it using the proposed methods.}
    \label{tab:results}
    \begin{tabular}{l||p{3.1cm}|p{3.1cm}||p{3.1cm}|p{3.1cm}}
                    & \multicolumn{2}{c||}{Entanglement Assertions}                                                             & \multicolumn{2}{c}{Equality Assertions} \\
    Algorithm Class & Moving & Adding and moving & Moving & Adding and moving \\ \hline
Deutsch-Jozsa                 & N/A                     & N/A                       & $15.6\%           \pm 12.2\%$           & $61.1\% \pm 27.5\%$           \\
GHZ State Preparation         & $53.3\% \pm 11.6\%$           & $58.6\% \pm \phantom{0}8.4\%$   & $\phantom{0}0.0\% \pm  \phantom{0}0.0\%$        & $\phantom{0}0.0\%  \pm  \phantom{0}0.0\% $                  \\
Graph State Preparation       & $23.9\% \pm 19.4\%$           & $49.2\% \pm 19.6\%$             & $14.8\%           \pm \phantom{0}6.0\%$ & $14.8\% \pm \phantom{0}6.0\%$ \\
Grover Search                 & $16.7\% \pm 12.1\%$           & $16.7\% \pm 12.1\%$             & $\phantom{0}2.7\% \pm \phantom{0}2.8\%$ & $20.1\% \pm 25.1\%$           \\
Quantum Fourier Transform     & $10.0\% \pm \phantom{0}1.0\%$ & $15.0\% \pm \phantom{0}5.8\%$   & $12.0\%           \pm 12.4\%$           & $12.0\% \pm 12.4\%$           \\
Quantum Phase Estimation      & N/A                     & N/A                       & $16.1\%           \pm 18.7\%$           & $44.0\% \pm 30.5\%$           \\
    \end{tabular}
    \redactedInstead{}{\vspace{-0.1cm}}
\end{table*}

To evaluate the improvements obtained from employing the proposed assertion refinement methods, 
all proposed methods have been implemented as \redactedInstead{part of an open-source debugging library available at \href{https://github.com/cda-tum/mqt-debugger}{https://github.com/cda-tum/mqt-debugger}.}{a C++ library that will be made open-source after the review process.}
Based on the resulting implementation, thorough evaluations have been conducted whose results are summarized in this section.

\subsection{Experimental Setup}

To generate a set of benchmarks, we used
6 representative quantum algorithms from the MQT Bench library~\cite{quetschlich2023mqtbench} each with 5 instances between 4 and 8 qubits as well as between 10 and 500 lines of code.
For each of the programs, we added appropriate assertions to the end of the circuits\footnote{Due to space limitations and because both techniques proposed in
\autoref{sec:methods} are applicable to them, we only focus on entanglement and equality assertions in the following. However, we also conducted evaluations based on superposition assertions, whose results follow a similar trend as those for the entanglement assertions. All results \redactedInstead{are}{will be made} available in the open-source implementation.}.
On the one hand, we added assertions to check the equality of the final state of the program with a reference state, which had been generated from classical simulations of the respective circuits.
On the other hand, we added entanglement assertions to check the entanglement of qubits at the end of the program based on an analysis of its final state.
As some of the programs do not result in an entangled state, adding entanglement assertions to the end of these programs is not applicable, which is why we only added equality assertions to these programs.
This leaves a total of $19$ test programs for entanglement assertions and $30$ for equality assertions.
These generated programs act as the ground truth.

To test the effectiveness of the proposed methods, we created faulty versions of the benchmark programs by randomly modifying single instructions.
For single-qubit gates, the instruction has been removed.
For multi-qubit gates, the first and last qubits involved in the gate have been swapped.
Repeating this process ten times and removing instances where the modifications did not alter the program's state (such as when swapping control and traget of a CZ gate), we obtained a total of $87$ and $194$ erroneous programs for entanglement and equality assertions, respectively.

\subsection{Evaluation Results}

\autoref{tab:results} summarizes the results on the effectiveness of the proposed methods for moving and adding assertions in reducing the number of lines that have to be inspected for errors in quantum programs.
Each row lists the relative reduction in the number of instructions that need to be considered between failing assertions and the error they detect.
The individual columns represent the results for moving assertions (as proposed in \autoref{sec:moving}) as well as the combination with adding new assertions (as proposed in \autoref{sec:adding}), respectively applied to the entanglement and equality assertions of the tested quantum programs.
Results are averaged for all instances of a particular quantum algorithm with the purpose of identifying trends in the performance of the proposed methods based on the characteristics of the considered quantum programs.
Cases where no entanglement assertions are applicable are marked in the table.
All results were obtained with negligible time overhead with respect to the number of instructions in the programs.

\subsection{Discussion}

Looking at the results obtained for \emph{entanglement assertions}, one can clearly see the significant improvements achieved by the proposed methods, especially for programs that require the entanglement of multiple qubits.
For \emph{GHZ State Preparation}, more than half of the lines between a failing assertion and the introduced error can already be neglected after just moving the assertions.
When additionally adding new entanglement assertions, this improvement is increased to an average of~$58.6\%$.
For \emph{Graph State Preparation}, the improvement from the creation of new assertions is even more significant as it pushed the reduction in the number of lines from $23.9\%$ to~$49.2\%$.

For \emph{equality assertions}, even more promising improvements can be observed when adding \emph{and} moving assertions.
This is because the commutation rules employed to move equality assertions are stricter than those for entanglement assertions.
Thus, the proposed methods for equality assertions perform particularly well when quantum programs do not require complex entanglement structures over multiple qubits, as corresponding assertions can be separated more easily in this case.
As a demonstration of that, the number of lines to be considered in the \emph{Deutsch-Jozsa} programs %
can be reduced by an average of $61.1\%$.

Notably, the result show that the proposed methods complement each other well.
In many cases, if one of the methods only provides limited improvements, the other method can be applied to achieve a more significant result.
Overall, this evaluation confirms that the proposed methods can significantly improve the otherwise tedious and time-consuming manual debugging effort by
greatly reducing the number of lines that have to be inspected for errors in negligible runtime.
In a large number of cases, more than half of the lines that would otherwise have
to be considered for manual debugging can be neglected.

\redactedInstead{}{\vfill}

\section{Conclusion}\label{sec:conclusion}

In this work we proposed several methods for the automated refinement of assertions in quantum programs,
 implemented as part of an open-source framework available at \redacted{\href{https://github.com/cda-tum/mqt-debugger}{https://github.com/cda-tum/mqt-debugger}}.
Given a set of initial assertions, the proposed methods employ different commutation rules to \emph{move} them to
better positions in the program. Additionally, by analyzing the provided assertions as well as the
program's structure, we proposed strategies to \emph{add} new assertions that can inspect the program's state
in more detail. We evaluate the proposed methods on a set of quantum programs with known errors, showing
that the proposed methods can significantly improve the quality of assertions in these programs. 
This allows developers to take advantage of qualitative assertions automatically, shielding them
from an otherwise tedious and time-consuming manual task.

\vfill
\redactedInstead{
\section*{Acknowledgments}
This work received funding from the European Research Council (ERC) under the European Union’s Horizon 2020 research and innovation program (grant agreement No. 101001318), was part of the Munich Quantum Valley, which is supported by the Bavarian state government with funds from the Hightech Agenda Bayern Plus, and has been supported by the BMWK on the basis of a decision by the German Bundestag through project QuaST, as well as by the BMK, BMDW, and the State of Upper Austria in the frame of the COMET program (managed by the FFG).
}
{}

\clearpage

\printbibliography

\end{document}